\documentclass[10pt,letterpaper,american]{article}
\usepackage[latin9]{inputenc}
\usepackage{color}
\usepackage{babel}
\usepackage{amsmath}
\usepackage{amssymb}
\usepackage{graphicx}
\usepackage[bookmarks=false,
 breaklinks=false,pdfborder={0 0 1},backref=false,colorlinks=true]
 {hyperref}
\hypersetup{
 citecolor=blue,urlcolor=blue}

\makeatletter

\pdfpageheight\paperheight
\pdfpagewidth\paperwidth

\DeclareTextSymbolDefault{\textquotedbl}{T1}

\usepackage{osameet3}%% use version 3 for proper copyright statement
\usepackage{capt-of}
\usepackage{booktabs}
\usepackage{setspace}

\usepackage{cite}
\usepackage[printonlyused]{acronym}
\acrodef{DM}{distribution matcher}
\acrodef{Hi-DM}{hierarchical DM}
\acrodef{PS}{probabilistic shaping}
\acrodef{PAS}{probabilistic amplitude shaping}
\acrodef{AWGN}{additive white Gaussian noise}
\acrodef{QAM}{quadrature amplitude modulation}
\acrodef{FEC}{forward error correction}
\acrodef{PAM}{pulse amplitude modulation}
\acrodef{LUT}{look up table}
\acrodef{MB}{Maxwell-Boltzmann}
\acrodef{ESS}{enumerative sphere shaping}
\acrodef{CCDM}{constant composition distribution matching}
\acrodef{IR}{information rate}
\acrodef{SNR}{signal to noise ratio}
\acrodef{PPM}{pulse position modulation}
\acrodef{invDM}{inverse DM}
\acrodef{TX}{transmitter}
\acrodef{RX}{receiver}
\acrodef{BER}{bit error rate}
\acrodef{SER}{symbol error rate}

\let\oldbibliography\thebibliography
\renewcommand{\thebibliography}[1]{%
  \oldbibliography{#1}%
  \footnotesize}

\makeatother

\begin{document}
\title{ML-Enhanced Digital Backpropagation for Long-Reach Single-Span Systems\vspace*{-2ex}
}
\author{D. Cellini\textsuperscript{1,{*}}, S. Civelli\textsuperscript{2,1}
and M. Secondini\textsuperscript{1,3}}
\address{\textsuperscript{1}TECIP, Scuola Superiore Sant'Anna, Via G. Moruzzi
1, 56124 Pisa, Italy (dario.cellini@santannapisa.it)\\
\textsuperscript{2}CNR-IEIIT, Via G. Caruso 16, 56122 Pisa, Italy\\
\textsuperscript{3}PNTLab, CNIT, Via G. Moruzzi 1, 56124 Pisa, Italy}

\maketitle
%% Uncomment the following line to override copyright year from the default current year.
\copyrightyear{2025}\vspace*{-3ex}

\begin{abstract}
We propose a digital backpropagation method that employs machine-learning-aided
joint optimization of dispersion step lengths and nonlinear phase
rotation filters within an FFT-based enhanced split-step Fourier structure,
achieving improved accuracy at low computational complexity.\vspace*{-5ex}
\end{abstract}

\section{Introduction}

\vspace*{-1ex}
Digital backpropagation (DBP) compensates for linear and nonlinear
fiber impairments in coherent optical systems\cite{Ip2008}. The
split-step Fourier method (SSFM) is a traditional solution for DBP
which alternates linear and nonlinear steps to accurately compensate
for intra-channel nonlinear interference. However, the large number
of required steps leads to high computational complexity, making the
SSFM impractical for real-world applications. To address this, many
low-complexity DBP techniques have been developed\cite{Rafique2011,SecondiniECOC,Civelli2025,Haeger2021}.
Two main approaches are particularly relevant to this work. The first
is the enhanced SSFM (ESSFM)\cite{SecondiniECOC,Civelli2025}, which
preserves the frequency-domain implementation of the linear step typical
of the conventional SSFM---based on FFT/IFFT pairs and the overlap-and-save
technique---and targets a small number of highly accurate steps.
ESSFM relies on a perturbative refinement of the nonlinear step to
increase the per-step accuracy, allowing to decrease the required
number of steps and, consequently, the number of FFTs and the overall
computational complexity. The second approach is learned DBP (LDBP)\cite{Haeger2021},
which performs the linear step in the time domain using FIR filters
with a large number of simple steps. LDBP exploits the layered structure
of SSFM, analogous to a deep neural network, and uses machine learning
(ML) to jointly optimize FIR coefficients across all steps. However,
the complexity of the FIR-based implementation increases linearly
with the total channel memory, hence quadratically with the symbol
rate, making LDBP less suitable for higher data rates. In contrast,
ESSFM, operating in the frequency domain, scales more favorably (logarithmically)
with channel memory. Nevertheless, ESSFM exhibits its own limitations:
it potentially involves a relatively large number of parameters, whose
optimization through conventional numerical techniques requires the
introduction of some constraints to limit the degrees of freedom---typically
by enforcing fixed step sizes (e.g., uniform) and identical coefficients
in the calculation of the nonlinear phase rotation (NLPR) at each
step\cite{Civelli2025}. These constraints may limit the ESSFM efficiency
in some scenarios. In this work, we combine the FFT-based structure
of ESSFM with ML-based joint optimization of all parameters, improving
the performance--complexity trade-off. We evaluate the proposed learned
ESSFM (L-ESSFM) in a wavelength division multiplexing (WDM) system
with 93-GBd channels over a long, unrepeatered single-span (170\,km)
link---an operating scenario of growing relevance for data-center
interconnects.\vspace*{-2ex}

\section{Learned ESSFM}

\vspace*{-1ex}
  In the SSFM, the fiber link is divided into $N_{s}$ steps (segments
of equal length), each comprising a linear step, which compensates
for group velocity dispersion (GVD), and a nonlinear step, which accounts
for Kerr nonlinearity through a NLPR. The ESSFM improves the nonlinear
step by modifying the NLPR to capture the interaction between dispersion
and nonlinearity. In this approach the NLPR is obtained by filtering
the signal intensity with a numerically optimized filter, equal for
all steps\cite{SecondiniECOC}. The CB-ESSFM further improves the
ESSFM by using subband processing and optimizing the splitting ratio
for the linear step\cite{Civelli2025}. The L-ESSFM---sketched in
Fig. \ref{fig:1}(a)---is a ML-aided parameterized variant of the
ESSFM, and is structured as a deep, feed-forward neural network (NN)
where each layer corresponds to a single step of the backpropagation
process. Both linear and nonlinear steps are implemented as learnable
functions within the NN, enabling the optimization of each step independently.
All filtering operations are implemented in frequency domain via FFT/IFFT
pairs and overlap-and-save. The trainable parameters, highlighted
in red in Fig. \ref{fig:1}(a), consist of the length $L_{i}$ of
each linear step ($i=0,\,...,\,N_{s}$) and the impulse response $\mathbf{c}_{i}$
of each NLPR ($i=1,\,...,\,N_{s}$), where the NLPR filter is assumed
symmetric ($2N_{c}+1$ real coefficients) and $\mathbf{c}_{i}$ contains
only the causal part of the response ($N_{c}+1$ trainable coefficients).
The GVD transfer function is $H_{\text{GVD}}(L_{i})=\exp(-j2\pi^{2}\beta_{2}f_{k}^{2}L_{i})$,
where $f_{k}$ is the $k$-th frequency component after FFT and $\beta_{2}$
the dispersion coefficient. The NLPR filter transfer function, $\mathit{\mathbf{H}}_{\text{NLPR}}(\mathbf{c}_{i})$,
is obtained via real-valued FFT (RFFT) of the symmetric impulse response,
zero-padded to the full FFT block length $N$. For comparison, the
ESSFM with optimized splitting ratio (i.e., the CB-ESSFM with a single
band) is equivalent to the L-ESSFM structure with identical coefficients
$\mathbf{c}_{i}$ for all steps, $L_{i}=L/N_{s}$ for $i\geq1$, where
$L$ is the link length, and $L_{N_{\text{s}}}=L-L_{0}$. Therefore,
once trained (off-line), the (real-time) computational complexity
of the L-ESSFM is the same as that of the CB-ESSFM with a single band\cite[Eq. (29)]{Civelli2025}.
The overall NN structure includes matched filtering, symbol-time sampling,
and mean phase removal (MPR) at the output. The L-ESSFM is trained
using supervised learning, with the training dataset generated by
drawing many blocks of i.i.d. complex Gaussian symbols. After fiber
propagation, received samples serve as NN inputs and transmitted symbols
as desired outputs. The GVD lengths are initialized to divide the
link into segments with comparable nonlinear effects, and the NLPR
filters as instantaneous nonlinearities. \vspace*{-2ex}

\section{System setup and performance}

\vspace*{-1ex}
\begin{figure}
\centering\begin{minipage}{0.51\textwidth}\centering\includegraphics[width=1\textwidth]{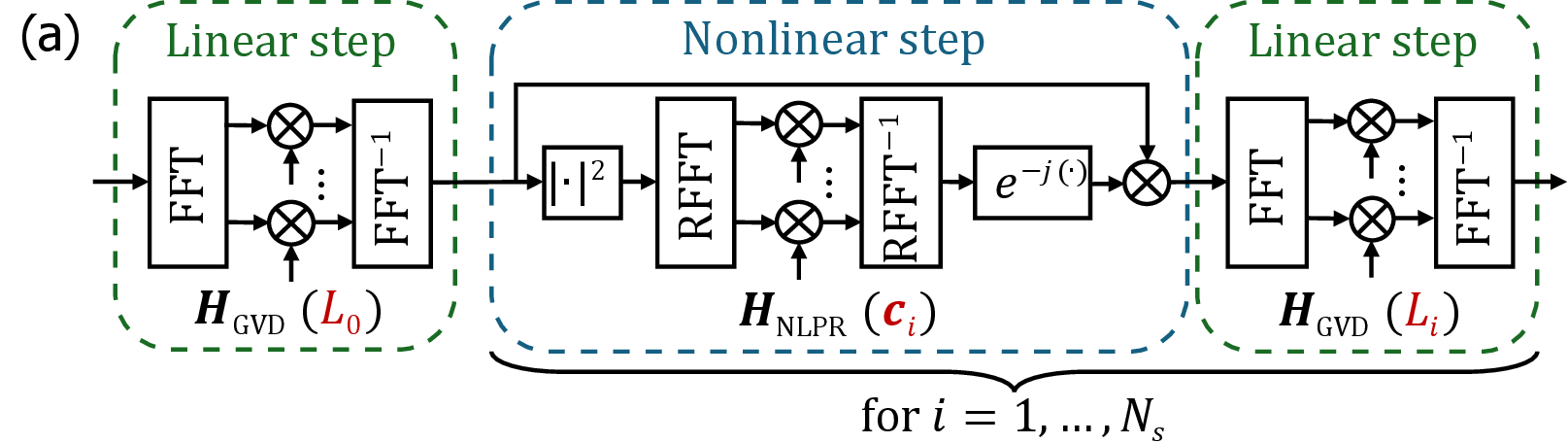}\caption{\label{fig:1}(a) L-ESSFM architecture with the\\learnable parameters
in red;\\(b) system performance vs. complexity in real\\multiplications
per 2D symbol.}
\end{minipage}\hfill\begin{minipage}{0.49\textwidth}\centering\includegraphics[width=1.04\textwidth]{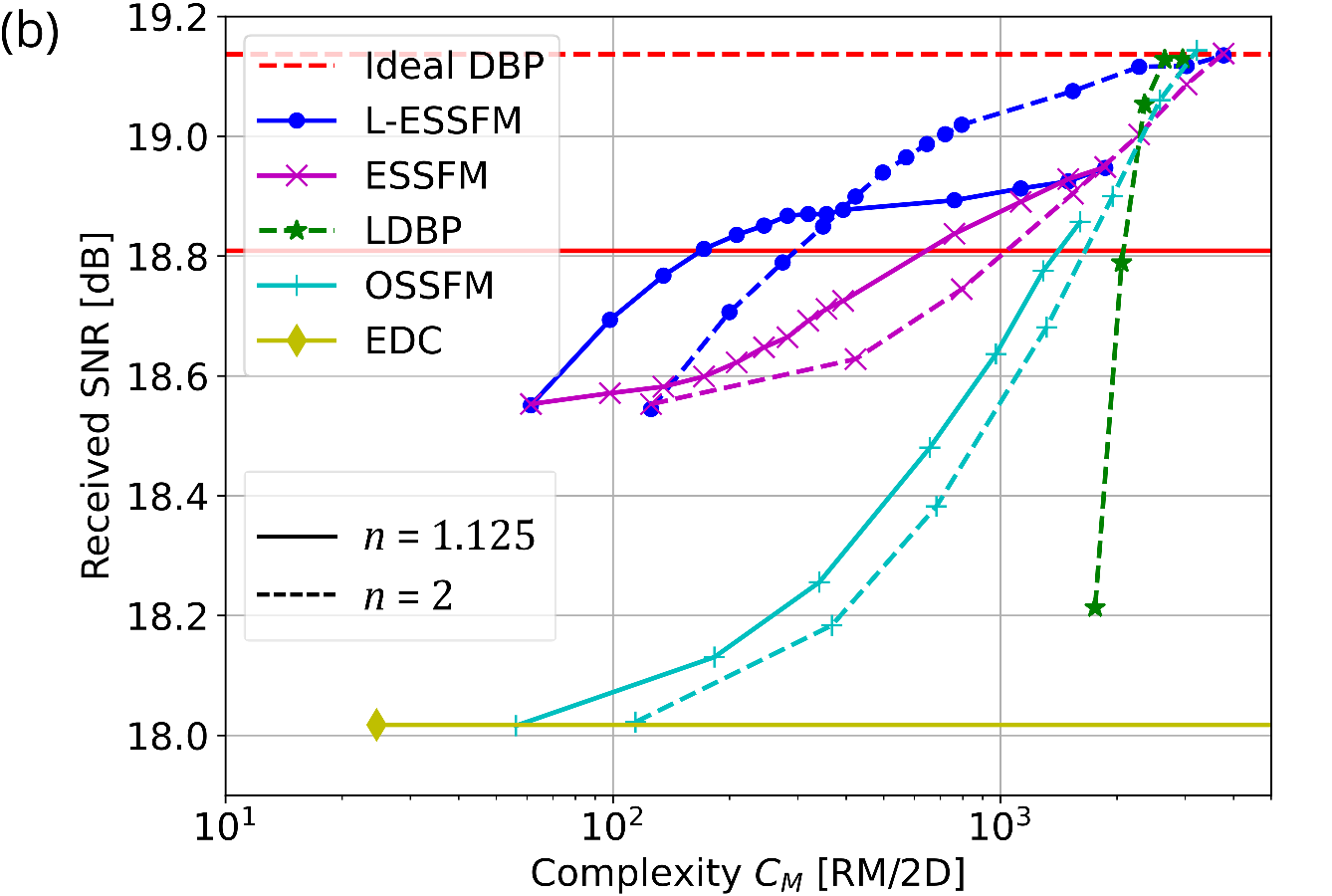}\end{minipage}

\vspace*{-3ex}
\end{figure}
System performance is evaluated via numerical simulations. The WDM
signal comprises 5$\times$93\,GBd channels (100\,GHz spacing),
dual polarization 64-QAM symbols, shaped with root-raised cosine pulse
(roll-off 0.05), transmitted over a 170\,km-long span of SMF (attenuation
0.2\,dB/km, dispersion 17\,ps/nm/km, and Kerr parameter 1.27\,W\textsuperscript{-1}km\textsuperscript{-1}).
At the receiver, an EDFA (noise figure 4.5\,dB) pre-amplifies the
signal, the central channel is demultiplexed, and DBP or EDC is applied,
followed by matched filtering, symbol sampling, and MPR. Fig. \ref{fig:1}(b)
shows the maximum received SNR for different DBP techniques and EDC
as a function of the number of real multiplications per complex symbol
(RM/2D). The optimized SSFM (OSSFM) refers to an SSFM with a numerically
optimized nonlinear parameter, equivalent to ESSFM with a single tunable
coefficient ($N_{c}=0$)\cite{Civelli2025}. DBP is performed with
$n=1.125$ samples/symbol (solid, only for FFT-based DBP methods)
or $n=2$ samples/symbol (dashed). The performance of ideal ($N_{s}\rightarrow\infty)$
SSFM is also shown as a benchmark (horizontal red lines). EDC, OSSFM,
ESSFM, and L-ESSFM use overlap-and-save. ESSFM is implemented with
optimized splitting ratio\cite{Civelli2025}. L-ESSFM uses $N_{c}=65$
trainable coefficients per step up to $N_{s}=10$, then gradually
decreased.   Complexity is evaluated as in\cite{Civelli2025}
(L-ESSFM and ESSFM being equivalent). EDC complexity is fixed, while
OSSFM, ESSFM, and L-ESSFM complexity varies with the number of steps.
LDBP is implemented in time domain with complexity evaluated as in\cite{Haeger2021},
its complexity depends mainly on the total number of FIR coefficients.
Therefore, we set $N_{s}=50$ and varied complexity by changing FIR
filter lengths (pruning). L-ESSFM outperforms ESSFM except at $N_{s}=1$
and $N_{s}\rightarrow\infty$ where they converge, confirming the
advantage of ML-based joint optimization. L-ESSFM with $n=1.125$
offers best low-complexity performance: 0.8\,dB gain over EDC with
172\,RM/2D ($N_{s}=4$), four times less than ESSFM (761\,RM/2D);
higher performance is achievable with $n=2$ at increased complexity.
LDBP achieves similar gains but requires more than 2000~RM/2D, confirming
less favorable scaling at high symbol rates. \vspace*{-5ex}

\section{Conclusions}

\vspace*{-1ex}
We proposed a novel low-complexity method for DBP, the L-ESSFM, which
leverages ML to improve each ESSFM step. The approach enables joint
optimization of all the GVD lengths and NLPR filters, achieving higher
accuracy for the same complexity. In a $5\times93$ GBd WDM system
over a $170$ km single-span link, L-ESSFM offers the best performance--complexity
trade-off. L-ESSFM with 4 steps (172\,RM/2D) provides $0.8$\,dB
gain over EDC, ESSFM requires four times and LDBP ten times higher
complexity for the same gain. The proposed approach can be readily
applied to long-haul links with inline amplification and further enhanced
by subband processing.\vspace*{-2ex}

\section*{Acknowledgments}

\vspace*{-1ex}
\footnotesize This work was partially supported by the European Union
under the Italian National Recovery and Resilience Plan of NextGenerationEU
(PE00000001 - program \textquotedbl RESTART\textquotedbl ). We thank
Christian H\"{a}ger from Chalmers University for helping us with
the source code of LDBP.\setstretch{0.9}\vspace*{-2ex}

\bibliographystyle{osajnl}

\end{document}